\documentclass[twocolumn]{jpsj3}

\usepackage{graphicx}

\title{Topological Hall Effect in PrSb$_2$}

\author{Shingo Araki$^{1,2}$\thanks{araki@science.okayama-u.ac.jp},
	Hinata Izumida$^2$, Kazuto Akiba$^{1,2}$\thanks{Present address: Faculty of Science and Engineering, Iwate University},
	Tatsuo~C.~Kobayashi$^{1,2}$, Takashi Kambe$^{1,2}$}
\inst{$^1$Department of Physics, Okayama University, Okayama, 700-8530, Japan \\
	$^2$ Graduate School of Environmental, Life, Natural Science and Technology, Okayama University, Okayama, 700-8530, Japan}

\date{\today}

\abst{
	PrSb$_2$ exhibits a charge-density wave (CDW) transition at $T_\mathrm{CDW}=100$~K and antiferromagnetic (AFM) ordering at $T_\mathrm{N}=5$~K at ambient pressure.
	Hall resistivity measurements revealed an anomalous feature within the AFM state, which was attributed to the topological Hall effect (THE), ruling out contributions from the ordinary and anomalous Hall effects.
	Significantly, the THE anomaly diminished with increasing pressure and almost vanished as the CDW transition was suppressed near the critical pressure $P_c=1.0$~GPa.
	These findings suggest a relationship between the CDW order and the appearance of the THE in PrSb$_2$.
}

\begin{document}
	\maketitle
		
	The topological Hall effect (THE) arises from the exchange coupling between the spins of conduction electrons and local magnetic moments as these electrons traverse a spatially varying spin texture.
	This phenomenon can be theoretically explained using a fictitious magnetic field derived from the Berry curvature in real space.
	Early THE observations were primarily confined to chiral magnetic systems, particularly MnSi~\cite{10.1103/physrevlett.102.186601,10.1103/physrevlett.102.186602}.
	In MnSi, the zero-field magnetic structure is a long-pitch helix driven by the Dzyaloshinsky-Moriya interaction.
	Applying an external magnetic field induces the A phase, in which a skyrmion lattice is generated through the superposition of multiple spin helices modulated along distinct orientations~\cite{10.1126/science.1166767}.
	The THE emerges within the A-phase.
	Furthermore, a comparable phase diagram and the THE have been reported in the chiral compound EuPtSi~\cite{10.7566/jpsj.87.023701,10.7566/jpsj.88.013702}.
	More recently, the THE has been observed in centrosymmetric $f$-electron systems, where spin textures are formed via intrinsic mechanisms~\cite{10.1126/science.aau0968,10.1038/s41467-019-13675-4,10.1038/s41565-020-0684-7,10.1038/s41567-024-02445-9,10.1103/physrevb.103.l020405}.
	For example, geometric frustration is crucial in the triangular lattice system Gd$_2$PdSi$_3$~\cite{10.1126/science.aau0968} and the breathing Kagom\'e lattice system Gd$_3$Ru$_4$Al$_{12}$~\cite{10.1038/s41467-019-13675-4}.
	In contrast, spin textures in GdRu$_2$Ge$_2$ arise from the competition between Ruderman--Kittel--Kasuya--Yosida interactions~\cite{10.1038/s41567-024-02445-9}.
	The THE has also been observed for EuAl$_4$~\cite{10.1103/physrevb.103.l020405}.
	To date, the THE in $f$-electron systems has only been observed in spin-only systems characterized by large magnetic moments and the absence of orbital angular momentum (e.g., Eu$^{2+}$ and Gd$^{3+}$ with $S=7/2$ and $L=0$).
	
	Assuming a small Hall angle, the Hall resistivity $\rho_{yx}$ in magnetic systems consists of three components~\cite{10.1103/physrevb.75.172403,10.1103/physrevlett.106.156603}:
	\begin{equation}
		\rho_{yx} = \rho_{yx}^\mathrm{N}+\rho_{yx}^\mathrm{A}+\rho_{yx}^\mathrm{T} \, .
		\label{eq1}
	\end{equation}
	Here, $ \rho_{yx}^\mathrm{N}$ represents the ordinary Hall effect (OHE), which originates from the Lorentz force under a transverse magnetic field $B_z$.
	Although it is often simplified to $R_0B$ (with $R_0 = -1/ne$, where $n$ and $e$ denote the carrier density and elementary charge, respectively), $R_0$ is not constant in most real materials, resulting in a nonlinear $B$ dependence of $ \rho_{yx}^\mathrm{N}$.
	The anomalous Hall effect (AHE) ($\rho_{yx}^\mathrm{A}$) scales with the magnetization $M_z$.
	Although the AHE was first observed in ferromagnets~\cite{10.1103/physrev.34.1466,10.1103/physrev.80.688}, it is now widely reported in antiferromagnetic or ferrimagnetic compounds that exhibit symmetry breaking analogous to ferromagnetic ordering~\cite{10.1038/nature15723,10.1103/physrevapplied.5.064009,10.1038/s41535-023-00587-2,10.1103/physrevresearch.2.043090,10.1103/physrevlett.133.106301}.
	Its origin involves both intrinsic (momentum space Berry curvature, $\propto\rho^2$~\cite{10.1103/revmodphys.82.1539}) and extrinsic mechanisms, including skew scattering~\cite{10.1016/s0031-8914(55)92596-9,10.1016/s0031-8914(58)93541-9} and side-jump scattering~\cite{10.1103/physrevb.2.4559}.
	The extrinsic contribution scales with residual resistivity $\rho_0$~\cite{10.1103/physrevlett.103.087206}.
	THE is the third component $\rho_{yx}^\mathrm{T}$.
	To experimentally isolate $\rho_{yx}^\mathrm{T}$, both the OHE and AHE components must be subtracted from the total measured $ \rho_{yx}$.
	
	Light rare-earth diantimonides, RSb$_2$ (R =La--Nd, Sm), exhibit various phenomena, including superconductivity in LaSb$_2$~\cite{10.1103/physrevb.83.174520} and pressure-induced superconductivity in CeSb$_2$~\cite{10.1103/physrevlett.131.026001}.
	They crystallize in the non-symmorphic orthorhombic structure of the SmSb$_2$-type (space group No.64, $Cmce$)~\cite{10.1021/ic50077a014,10.1021/ic50055a017}, characterized by alternating layers of Sb and R-Sb stacked along the $c$ axis.
	Notably, although the overall crystal structure is centrosymmetric, the R site lacks local inversion symmetry.
	The lattice constants $a$ and $b$ are nearly identical.
	Consistent with this layered structure, the single crystals exhibit a plate-like morphology with the plate normal to the $c$ axis.
	Magnetically, all the compounds in this series, except for nonmagnetic LaSb$_2$, undergo antiferromagnetic ordering at low temperatures.
	Consistent with their crystal structures, the magnetic susceptibilities of CeSb$_2$ and PrSb$_2$ are highly anisotropic, with the easy and hard axes lying in the $ab$ plane and along the $c$ axis, respectively.
	Furthermore, when a magnetic field is applied within the $ab$ plane, successive metamagnetic transitions of approximately 10~kOe are observed in both CeSb$_2$\cite{10.1088/1674-1056/26/6/067102,10.1088/1361-648x/ab9bcd,10.1103/physrevb.57.13624} and PrSb$_2$~\cite{10.1103/physrevb.57.13624}.
	Recently, novel magnetic-field-induced switching of the magnetization easy axis at approximately 340~kOe was discovered in CeSb$_2$~\cite{10.7566/JPSJ.94.043702}.
	
	In PrSb$_2$, the magnetization within the $ab$ plane at 2~K exhibits two metamagnetic transitions at $H_1 = 4$~kOe and $H_2=12$~kOe, with apparent saturation at a critical field of $H_\mathrm{c}=17$~kOe~\cite{10.1103/physrevb.57.13624,10.1016/s0921-4526(99)00964-3}.
	Further studies under pressure confirm the presence of metamagnetic transitions up to 1.1~GPa~\cite{10.1016/s0921-4526(99)00964-3}, which exhibit sharp increases and decreases in the field dependence of the resistivity~\cite{10.1103/physrevb.57.13624,10.1016/s0921-4526(99)00964-3}.
	Additionally, the temperature-dependent resistivity reveals a hump-like anomaly around $T_\mathrm{CDW}=100$~K, which is attributed to a charge density wave (CDW) transition~\cite{10.1103/physrevb.57.13624}.
	Resistivity measurements under pressure indicate that $T_\mathrm{CDW}$ decreases with increasing pressure, becoming unresolvable around 1.0~GPa~\cite{10.1016/s0921-4526(97)00278-0}.
	Thus, PrSb$_2$ is suitable for exploring the interplay between CDW and antiferromagnetic order.
	Our current Hall-effect measurements on PrSb$_2$ revealed a distinctive Hall response under an applied magnetic field.

	\begin{figure}
		\includegraphics[width=9cm]{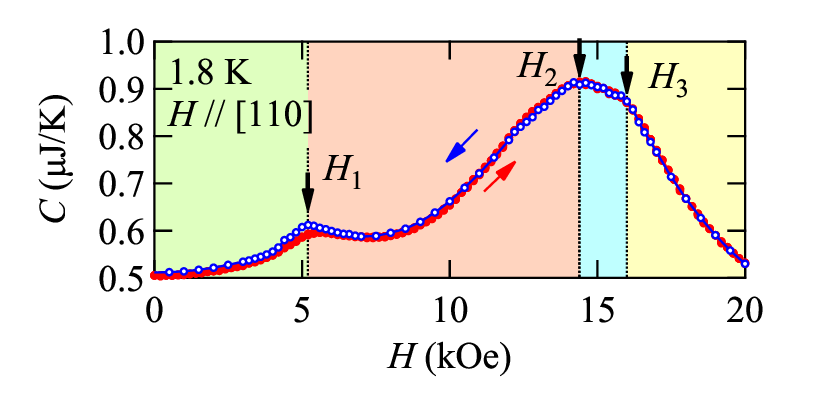}%
		\caption{(Color online) Magnetic field dependence of the specific heat $C$ in PrSb$_2$.}
		\label{CvsH}
	\end{figure}
	
	Single crystals of PrSb$_2$ were grown by using the Sb flux method~\cite{10.1103/physrevb.57.13624}.
	The Hall effect and specific heat were measured using a Physical Property Measurement System (PPMS, Quantum Design).
	For the Hall effect measurements, a four-wire cross configuration was used with the current almost along the $[\bar{1}10]$ direction, Hall voltage measured along the $[001]$ direction and magnetic field swept along the $[110]$ direction.
	The Hall resistance $R_{yx}$ and longitudinal resistance $R_{xx}$ were extracted from the antisymmetric and symmetric voltage components, respectively, with respect to the magnetic field.
	Subsequently, the Hall resistivity $\rho_{yx}$ was calculated from $R_{yx}$ and the sample thickness.
	A piston cylinder cell was used for the pressure-dependent Hall-effect experiments.
	The pressure was determined using the superconducting transition temperature of Pb~\cite{10.1088/0305-4608/11/3/010}, using Daphne 7373 as the pressure medium~\cite{10.1063/1.1148145}.
	The pressure-dependent resistivity was measured using an indenter cell~\cite{10.1063/1.2459512} with Daphne 7474 as the pressure medium~\cite{10.1063/1.2964117}.
	Ambient pressure magnetization measurements were performed using a SQUID-VSM (Quantum Design).
	
\begin{figure}
		\includegraphics[width=9cm]{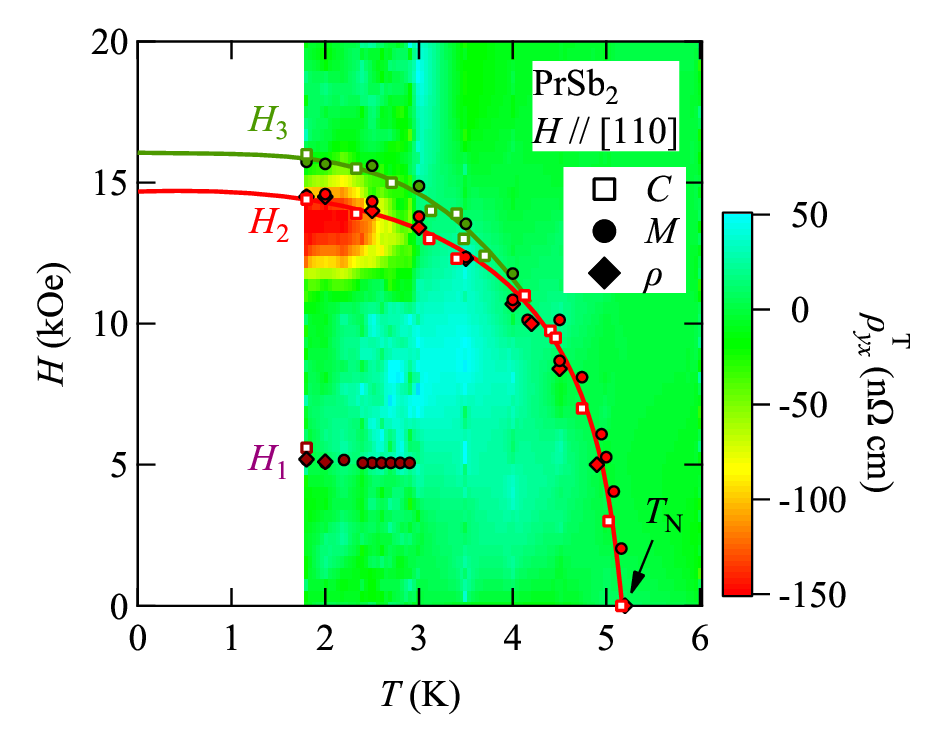}%
		\caption{(Color online) Phase diagram of PrSb$_2$ at ambient pressure determined from the specific heat $C$ (square), magnetization $M$ (circle), and resistivity $\rho_{xx}$ (diamond).
			The color background corresponds to the magnitude of $\rho_{yx}^\mathrm{T}$.
			\label{pd_0GPa}}
	\end{figure}

	Figure~\ref{CvsH} shows the magnetic field dependence of the specific heat $C$ measured at 1.8~K.
	The plot reveals distinct peaks at $H_1 = 5$~kOe and $H_2=14.4$~kOe, with a kink at $H_3=16$~kOe.
	Hysteresis is observed at approximately $H_1$.
	The temperature dependences of the critical fields ($H_1$, $H_2$, and $H_3$) are shown by the open squares in Fig.~\ref{pd_0GPa}.
	The fields $H_2$ and $H_3$ merge at higher temperatures, smoothly connecting to the N\'eel temperature, $T_\mathrm{N}$.
	The anomaly at $H_1$ broadens with increasing temperature and vanishes at approximately 3~K.

\begin{figure}
		\includegraphics[width=9cm]{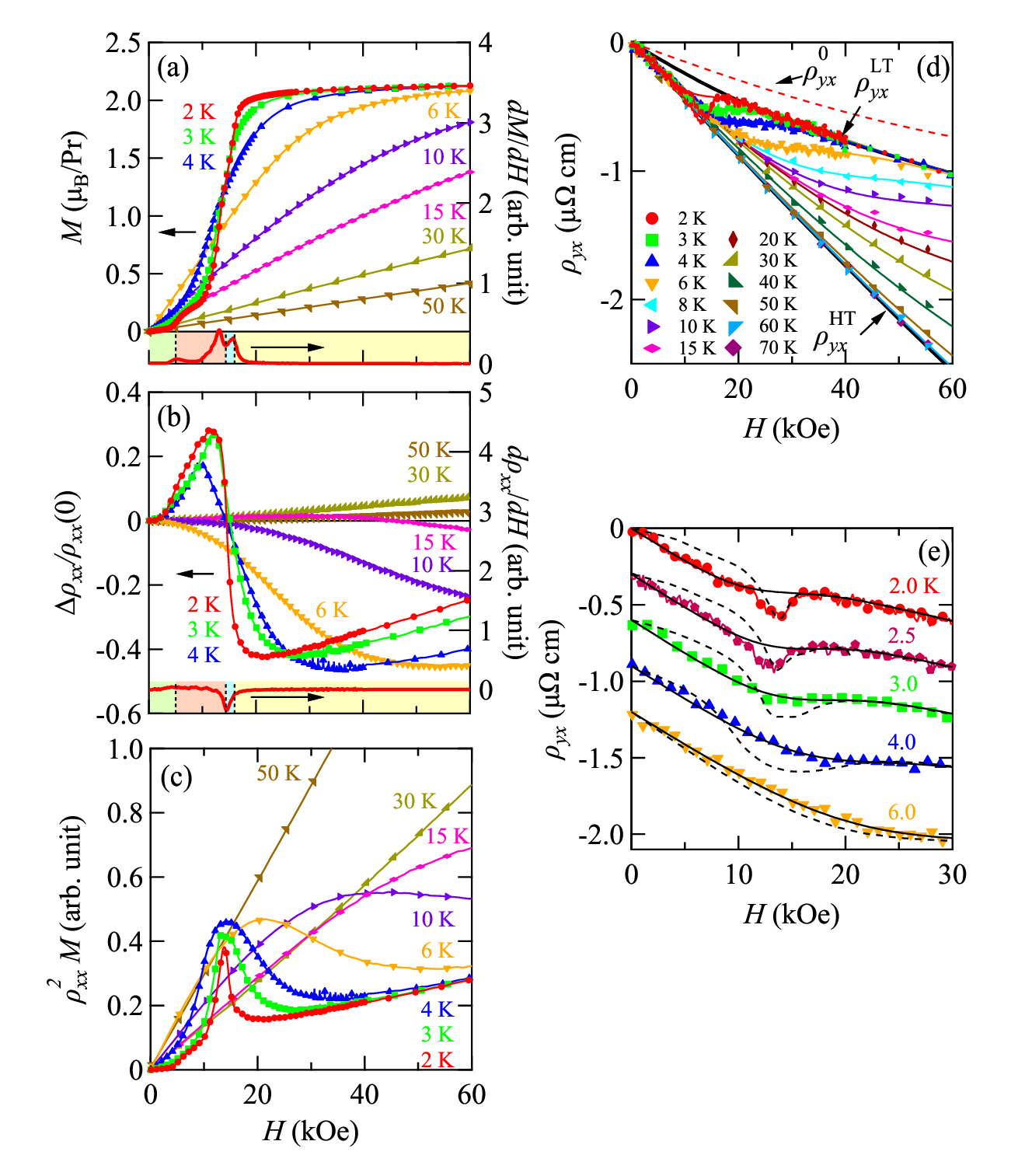}%
		\caption{(Color online) Magnetic field dependence of various properties in PrSb$_2$ with the magnetic field applied along the [110] direction.
			All data were acquired during a field-increasing sweep.
			(a) Magnetization ($M$) at various temperatures and its field derivative ($dM/dH$) at 2~K.
			(b) Normalized magnetoresistance ($\Delta\rho_{xx}/\rho_{xx}(0)$) at various temperatures and its field derivative ($d\rho_{xx}/dH$) at 2~K.
			(c) The product $\rho_{xx}^2 M$
			(d, e) Hall resistivity ($\rho_{yx}$) at various temperatures.
			For clarity, the curves in panel (e) are vertically offset, except for the 2.0 K.
			The dashed curves in panel (e) are the fits to the expression $\rho_{yx}^0+S_A\rho^2M$, while the solid curves are the calculated OHE contributions (see text for details).
			\label{0GPa_Hall}}
	\end{figure}

	Figure~\ref{0GPa_Hall}(a) shows the magnetic field dependence of the magnetization $M$ at various temperatures, along with the field derivative $dM/dH$ at 2~K.
	As reported previously~\cite{10.1103/physrevb.57.13624,10.1016/s0921-4526(99)00964-3}, multistep metamagnetic transitions occur at $H_1$ and approximately at $H_2$ at 2~K.
	The second metamagnetic transition exhibits a two-step behavior, which is clearly resolved in the field derivative $dM/dH$.
	Corroborating the specific heat data, the peaks in $dM/dH$ at 5~kOe and 16~kOe are consistent with $H_1$ and $H_3$ observed in $C$.
	However, no anomaly is detected in $C$ at 13.3~kOe, where $dM/dH$ exhibits an additional peak.
	Conversely, $dM/dH$ exhibits a minimum value at $H_2$.
	The temperature dependences of $H_2$ and $H_3$, as determined from $M$, are plotted in Fig.~\ref{pd_0GPa}.
	Figure~\ref{0GPa_Hall}(b) shows the magnetic field dependence of the normalized magnetoresistance ($\Delta\rho_{xx}/\rho_{xx}(0)=(\rho_{xx}(H)-\rho_{xx}(0))/\rho_{xx}(0)$) at various temperatures and its derivative ($d\rho_{xx}/dH$) at 2~K.
	The sharp increases and decreases in $\rho_{xx}$ at 2~K are directly correlated with the metamagnetic transitions at $H_1$ and $H_2$, respectively.
	The temperature dependences of $H_1$ and $H_2$ as determined from $\rho_{xx}$ are plotted in Fig.~\ref{pd_0GPa}.
	The critical fields determined from $C$, $M$, and $\rho$ are consistent with each other.
	
	As shown in Fig.~\ref{0GPa_Hall}(d), the Hall resistivity $\rho_{yx}$, down to approximately 10~K, is adequately described by the OHE, as calculated using a two-carrier model~\cite{10.1063/1.371187}.
	However, multicarrier models (including those with additional components) cannot fully account for the observed $\rho_{yx}$ behavior below this temperature.
	In particular, a strong dip emerges at 2~K within the antiferromagnetic phase below $H_3$.
	The AHE is a potential candidate for this low-temperature anomaly because it typically scales with $\rho^2 M$ (intrinsic) or $M$ (extrinsic), as illustrated in Fig.~\ref{0GPa_Hall}(c) and Fig.~\ref{0GPa_Hall}(a), respectively.
	Because the field dependence of $\rho_{yx}$ at 2~K resembles that of $\rho^2M$ but differs from that of $M$, we tested a fit to the expression $\rho_{yx}=\rho_{yx}^0 + S_A\rho^2 M$, where $\rho_{yx}^0$ was the OHE calculated using a two-carrier model (dashed curve in Fig.~\ref{0GPa_Hall}(d)).
	The results of this fit (dashed curves in Fig. ~\ref{0GPa_Hall}(e)) reproduce the dip structure in $\rho_{yx}$ at 2~K.
	However, this model fails to reproduce the overall $\rho_{yx}$ behavior at higher temperatures.
	Moreover, the model predicts a strong temperature dependence at low fields ($H<10$~kOe) (Fig.~\ref{0GPa_Hall}(c)), which is inconsistent with the experimental data and nearly temperature-independent in this region.
	This major inconsistency indicates that the AHE contribution is negligible compared to the observed $\rho_{yx}$ and that the magnetic field and temperature dependence of $\rho_{yx}$ is primarily governed by the OHE.
	
	Because a simple multicarrier model proved inadequate for reproducing the observed $\rho_{yx}$ across all the measured temperature and magnetic field regions, we adopted an empirical method to estimate the field dependence of the OHE.
	At low temperatures and high magnetic fields, $\rho_{yx}$ asymptotically approached a unique curve, which was denoted as $\rho_{yx}^\mathrm{LT}$.
	As the temperature decreased, the field dependence of $\rho_{yx}$ crossed from a high-temperature regime ($\rho_{yx}^\mathrm{HT}$) to a low-temperature high-field curve ($\rho_{yx}^\mathrm{LT}$).
	This empirical behavior is represented by the following expression:
	\begin{equation}
		\rho_{yx} = \rho_{yx}^\mathrm{HT} + \gamma(\rho_{yx}^\mathrm{LT}-\rho_{yx}^\mathrm{HT}).
		\label{eq2}
	\end{equation}
	Here, $\rho_{yx}^\mathrm{HT}$ and $\rho_{yx}^\mathrm{LT}$ are the two-carrier model estimates of $\rho_{yx}$ at 70~K and 1.8~K, respectively.
	The crossover function $\gamma$ is effectively described by a sigmoid function $\gamma = 1/[1+e^{(H_0-H)/r}]$, where $H_0$ and $r$ are fitting parameters.
	The solid curves in Figs. ~\ref{0GPa_Hall}(d) and \ref{0GPa_Hall}(e) show that the calculations using Eq.~(\ref{eq2}) accurately reproduce $\rho_{yx}$ at 3~K.
	However a significant discrepancy between te measured $\rho_{yx}$ and the calculated OHE arises below 2.5~K, specially around 13~kOe.
	This observation provides strong evidence of the presence of the THE.
	The background color shown in Fig.~\ref{pd_0GPa} represents the magnitude of the THE component ($\rho_{yx}^\mathrm{T}$) which exhibits a strong feature that smears above approximately 3~K, just below $H_2$.
	
	\begin{figure}
		\includegraphics[width=9cm]{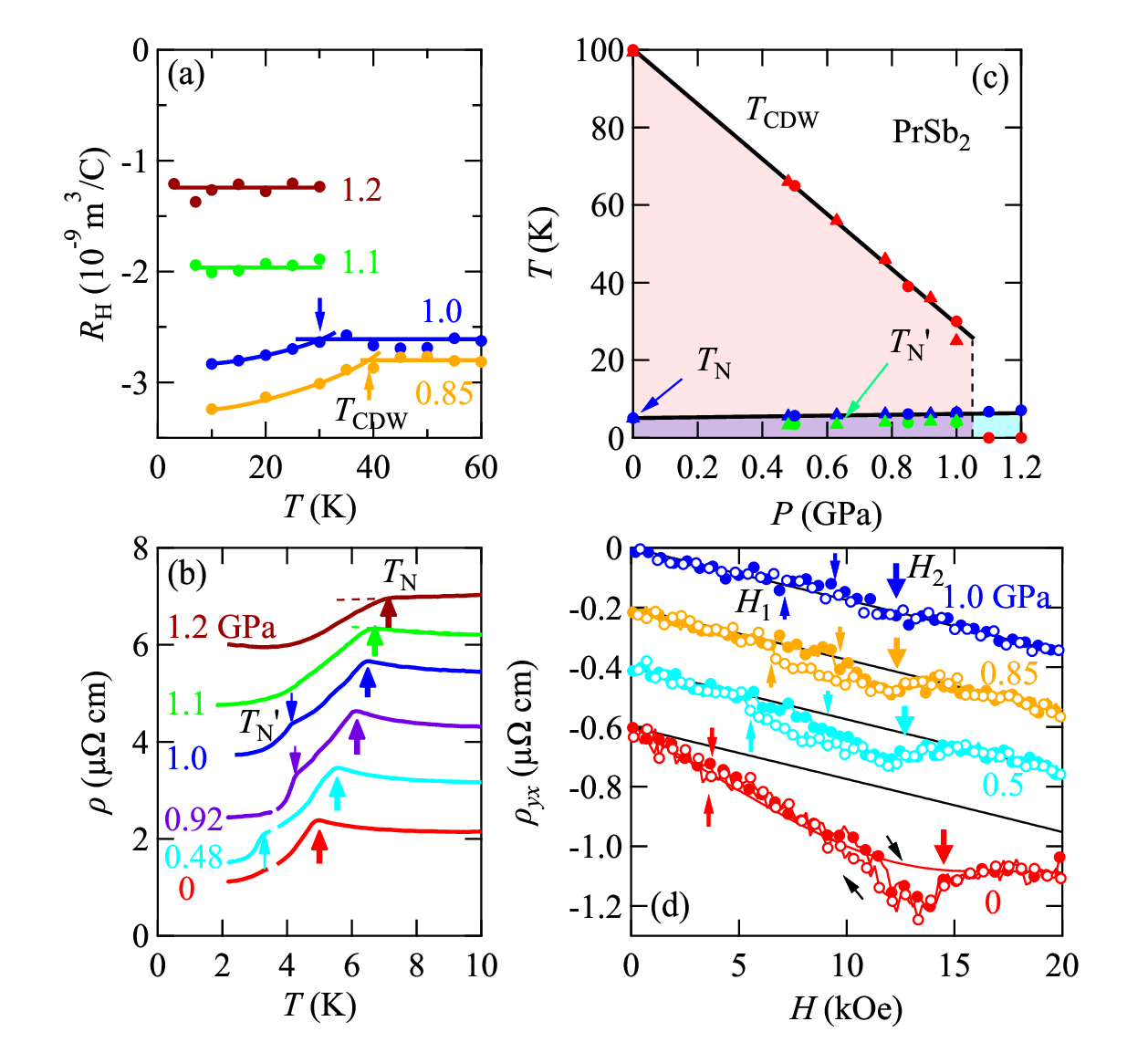}%
		\caption{
			(Color online) Pressure effects on the transport properties and the phase diagram of PrSb$_2$.
			(a) Temperature dependence of the Hall coefficient.
			(b) Low-temperature resistivity.
			(c) Temperature-pressure phase diagram, with data from Hall effect and resistivity measurements shown as circles and triangles, respectively.
			(d) Magnetic field dependence of the Hall resistivity $\rho_{yx}$ at 2.5~K under various pressures.
			The solid and open circles represent data acquired during field-increasing and field-decreasing sweeps, respectively.
			For clarity, all data except at 1.0~GPa have been successively shifted vertically.
			The black solid curve is the estimated OHE contribution from the $\rho_{yx}$ data at 1.0~GPa.
			The thick and thin arrows indicate the critical fields $H_1$ and $H_2$ determined from the $\rho_{xx}$, respectively.
			$H_1$ exhibits hysteresis, with the lower- and higher-field arrows corresponding to the field-decreasing and field-increasing sweeps, respectively.
			\label{fig4}}
	\end{figure}

	The results of the pressure study are described below.
	The CDW transition was identified by a hump-like anomaly in both the resistivity~\cite{10.1016/s0921-4526(97)00278-0} and Hall coefficient.
	The temperature dependence of the Hall coefficient, as shown in Fig. ~\ref{fig4}(a), indicates that $T_\mathrm{CDW}$ persists up to 1.0~GPa and disappears at 1.1~GPa.
	This contrasts with $T_\mathrm{N}$ derived from the resistivity data in Fig.~\ref{fig4}(b), which gradually increases with the pressure and remains above 1.1~GPa.
	The corresponding transition temperatures are shown in Fig. ~\ref{fig4}(c).
	The resistivity exhibits an additional low-temperature kink $T_\mathrm{N}'$ above 0.48~GPa.
	This anomaly is associated with the critical field $H_1$ and persists up to 1.0~GPa, as indicated by the green squares in Fig.~\ref{fig4}(c).
	The absence of a $T_\mathrm{N}'$ anomaly at ambient pressure was confirmed by $\rho$, $M$, and $C$ measurements.
	Figure~\ref{fig4}(d) shows the magnetic-field dependence of $\rho_{yx}$ at 2.5~K under various pressures.
	The data at 0.5~GPa and 0.85~GPa are consistent with this curve below $H_1$ and above $H_2$, suggesting that the OHE is pressure-independent within this range.
	Conversely, the OHE contribution at ambient pressure implies a pressure-dependent OHE within the 0--0.5~GPa.
	These changes, specifically the $T_\mathrm{N}'$ anomaly and pressure-dependent OHE, suggest a difference in the electronic and magnetic properties between 0~GPa and 0.5~GPa.
	We verified that at ambient pressure, the values of $\rho_{yx}$ measured before and after the pressure experiments on the same sample yielded identical results, thus ruling out sample damage during the measurements.
	
	The THE, which manifests as a dip anomaly in $\rho_{yx}$, was observed between $H_1$ and $H_2$ at 0.5~GPa and 0.85~GPa.
	These fields were determined from anomalies in $d\rho_{xx}/dH$.
	$\rho_{yx}^\mathrm{T}$ gradually weakened with increasing pressure and vanished at 1.0~GPa.
	The field region where $\rho_{yx}^T$ appeared at 0.5~GPa and 0.85~GPa differed from that at the ambient pressure, where $\rho_{yx}^\mathrm{T}$ occurred immediately below $H_2$.
	This difference is attributed to the skew scattering originating from magnetic fluctuations near the phase boundary~\cite{10.1126/sciadv.aap9962,10.1126/sciadv.abb6003,10.1038/s41467-020-20384-w,10.1103/physrevlett.125.137202}.
	The crystal symmetry, influenced by the CDW transition, may create conditions favorable for the formation of a nontrivial magnetic structure that induces the THE.
	
	Owing to the crystal field, magnetism in Pr$^{3+}$ systems is significantly influenced by single-ion anisotropy.
	This contrasts sharply with Eu$^{2+}$ and Gd$^{3+}$ compounds, which are spin-only systems where the THE has been reported.
	To the best of our knowledge, this paper reports the first observation of the THE in a Pr-based antiferromagnetic system.
	Despite this, a comprehensive understanding of the origin of THE in PrSb$_2$ remains challenging because of the lack of crystal structure data below $T_\mathrm{CDW}$ and detailed magnetic structure information.

	In conclusion, we observed a characteristic decrease in the Hall resistivity of PrSb$_2$, which was assigned as the THE.
	The disappearance of the THE under pressure simultaneously with the suppression of the CDW transition highlights the importance of the CDW in the emergence of the THE in PrSb$_2$.

	\begin{acknowledgment}
		We thank A. Miyake for fruitful discussion.
		This study was supported by JSPS KAKENHI (Grant No.~JP23H04868).
	\end{acknowledgment}

\end{document}